\documentclass{iaus}
\usepackage{graphicx}

\title[short title of paper] %% give here short title %%
{Radio Observations of the AGN and Gas in Low Surface Brightness Galaxies}

\author[short author list]   %% give here short author list %%
{M.Das$^1$, K.O'Neil$^2$, N.Kantharia$^3$, S.N.Vogel$^4$, S.S.McGaugh$^4$}

\affiliation{$^1$Raman Research Institute, Bangalore 560 080, India; 
email: mousumi@rri.res.in \break
$^2$NRAO, P.O. Box 2, Green Bank, WV 24944 \break
$^3$NCRA, TIFR, Post~Bag 3, Ganeshkhind, Pune-411007, India \break
$^4$University of Maryland, College Park, MD 20742  }

\pubyear{2006}
\volume{235}  %% insert here IAU Symposium No.
\pagerange{119}
\date{?? and in revised form ??}
\jname{Proceedings Title IAU Symposium}
\editors{F.Combes \& J. Palous, eds.}
\begin{document}

\maketitle

\begin{abstract}
LSB galaxies have low metallicities, diffuse stellar disks, and massive HI disks.
We have detected molecular gas in two giant LSB galaxies,
UGC~6614 and F568-6. A millimeter continuum source has been detected
in UGC~6614 as well. At centimeter wavelengths we have detected and mapped the 
continuum emission from the giant LSB galaxy 1300+0144. The emission is extended 
about the nucleus and is most likely originating from the AGN in the galaxy. 
The HI gas distribution and velocity 
field in 1300+0144 was also mapped. The HI disk extends well beyond the optical 
disk and appears lopsided in the intensity maps.

\end{abstract}

\vspace{-3mm}
\firstsection % if your document starts with a section,
\section{CO Detection and Millimeter Continuum Emission Observations}

We searched for CO(1--0) emission from seven LSB galaxies using the BIMA array. Three 
of the galaxies are large spirals; UGC~5709, UGC~6614 and F568-6 (Malin~2). The remaining 
four are relatively smaller galaxies with irregular disks; they are NGC~5585,
UGC~4115, UGC~5209 and F583-1. The BIMA maps suggested the presence of CO(1--0) emission
in two galaxies, UGC~6614 and F568-6. Using the IRAM 30m telescope we detected molecular
gas in the disks of both galaxies and in the nucleus of F568-6. These observations show that
molecular gas in LSB galaxies is present in their disks as well as their nuclei, but the 
distribution is localized over isolated regions and hence difficult to detect. We have also 
detected millimeter continuum emission from the nucleus of UGC~6614. The emission has a flat 
spectrum in the 1.4 to 110~GHz range and is from the AGN in the galaxy \cite{Das06}. 

\vspace{-5mm}
\section{Centimeter Observations of 1300+0144} 

1300+0144 (or PGC045080) is a giant LSB galaxy that is close to edge on and has a bright nucleus and
faint bar. We have detected and mapped the radio continuum emission at 1.4~GHz, 610~Mhz and 320~MHz
using the GMRT at Pune. The spectral index is -0.7 and the emission is extended about the nucleus. 
The emission is most likely due to the AGN and appears as lobes/jets; the core is not detected 
however and may be self absorbed. The HI gas distribution is asymmetric about the nucleus and 
appears lopsided. There is a dip in the HI distribution on the approaching side 
which is apparent in  
the HI map and spectrum. The disk rotation velocity flattens at $\sim200~$km~s$^{-1}$.

\begin{acknowledgments}
Observations with the BIMA millimeter-wave array are partially supported by NSF AST-0228974.
We are grateful to the members of the IRAM staff and  also to the GMRT staff for help in  
observations. This research has made use of NASA/ IPAC Infrared Science Archive,
HyperLeda database and the FIRST survey. 
\end{acknowledgments}

\end{document}